\documentclass[11pt,a4paper]{article}

\usepackage{amssymb}

\usepackage[dvips]{graphicx}

\usepackage{bm}

\begin{document}

\title{Manifestly ${\cal N}=2$ supersymmetric regularization for ${\cal N}=2$ supersymmetric field theories}

\author{
I.L.Buchbinder,\\
{\small{\em Department of Theoretical Physics, Tomsk State
Pedagogical University,}}\\
{\small{\em Tomsk 634061, Russia,}}\\
{\small{\em and}}\\
{\small{\em National Research Tomsk State University, Russia}}\\
\\
N.G.Pletnev,\\
{\small{\em Department of Theoretical Physics, Sobolev Institute of Mathematics,}}\\
{\small{\em 630090, Novosibirsk, Russia}},\\
{\small{\em and}}\\
{\small{\em National Research Novosibirsk State University, Russia}}\\
\\
K.V.Stepanyantz\\
{\small{\em Moscow State University, Physical
Faculty, Department  of Theoretical Physics,}}\\
{\small{\em 119991, Moscow, Russia}}}

\maketitle

\begin{abstract}
We formulate the higher covariant derivative regularization for
${\cal N}=2$ supersymmetric gauge theories in ${\cal N}=2$ harmonic
superspace. This regularization is constructed by adding the ${\cal
N}=2$ supersymmetric higher derivative term to the classical action
and inserting the ${\cal N}=2$ supersymmetric Pauli--Villars
determinants into the generating functional for removing one-loop
divergencies. Unlike all other regularization schemes in ${\cal
N}=2$ supersymmetric quantum field theory, this regularization
preserves by construction the manifest ${\cal N}=2$ supersymmetry at
all steps of calculating loop corrections to the effective action. Together with
${\cal N}=2$ supersymmetric background field method this
regularization allows to calculate quantum corrections without
breaking the manifest gauge symmetry and ${\cal N}=2$ supersymmetry.
Thus, we justify the assumption about existence of a regularization
preserving ${\cal N}=2$ supersymmetry, which is a key element of the
${\cal N}=2$ non-renormalization theorem. As a result, we give the
prove of the ${\cal N}=2$ non-renormalization theorem which does not
require any additional assumptions.
\end{abstract}

\unitlength=1cm

Keywords: higher derivative regularization, supersymmetry,
harmonic superspace.


\section{Introduction}
\hspace{\parindent}

The ${\cal N}=2$ non-renormalization theorem states that the global
${\cal N}=2$ supersymmetric gauge theories are finite beyond the
one-loop approximation. This theorem was first enunciated in \cite{Grisaru:1982zh}
for the ${\cal N}=2$ supersymmetric Yang--Mills (SYM) theory. The unconstrained ${\cal N}=2$
superfield formulation of the hypermultiplet was constructed in \cite{Howe:1982tm},
where it was used for proving the finiteness of the ${\cal N}=4$ SYM theory.
On its basis the detailed proof of the ${\cal N}=2$ non-renormalization theorem was
given in \cite{Howe:1983sr}. Using this theorem it is possible to obtain that
${\cal N}=2$ supersymmetric gauge theories are finite if their one-loop
$\beta$-function vanishes \cite{Howe:1983wj}. A proof of the ${\cal N}=2$ non-renormalization
theorem based on the harmonic superspace approach was given in \cite{Buchbinder:1997ib}. Deriving
the non-renormalization theorem one implicitly assumes existence of a
regularization which does not break the gauge symmetry and ${\cal
N}=2$ supersymmetry. However, a construction of a regularization
which satisfies these requirements is not evident
\cite{Jack:1997sr}. In particular, the standard dimensional
regularization breaks supersymmetry, and supersymmetric theories
are mostly regularized by using its special modification, which is
called the regularization by means of dimensional reduction
\cite{Siegel:1979wq}. However, the dimensional reduction is
inconsistent from the mathematical point of view
\cite{Siegel:1980qs}. In principle, it is possible to remove the
inconsistencies, but the price is the loss of manifest supersymmetry
\cite{Avdeev:1981vf}. As a consequence, supersymmetry can be broken
by quantum corrections in higher loops
\cite{Avdeev:1982np,Avdeev:1982xy}. In particular, the explicit
calculations made in \cite{Avdeev:1982np} and subsequently corrected
in \cite{Velizhanin:2008rw} show that for the ${\cal N}=2$
SYM theory supersymmetry is really
broken by quantum corrections in the three-loop approximation if the
regularization by means of dimensional reduction is used. This
implies that in this case the assumptions used in the proof of the
non-renormalization theorem are broken due to the loss of manifest
supersymmetry. Thus, the dimensional reduction cannot be considered
as a completely satisfactory regularization for supersymmetric
theories and the proof of the non-renormalization theorem contains a
hole. The purpose of this paper is to remove this hole and to
justify finally the ${\cal N}=2$ non-renormalization theorem.

We would like to pay attention that there exists a consistent
regularization convenient for using in gauge theories. It is called
the higher covariant derivative regularization
\cite{Slavnov:1971aw,Slavnov:1972sq}. For ${\cal N}=1$
supersymmetric gauge theories such a regularization can be
formulated in terms of ${\cal N}=1$ superfields
\cite{Krivoshchekov:1978xg,West:1985jx}, so that ${\cal N}=1$
supersymmetry is a manifest symmetry at all steps of quantum
calculations. This regularization appears to be very convenient for
explicit computing the quantum corrections (see, e.g.
\cite{Pimenov:2009hv,Stepanyantz:2011bz,Kazantsev:2014yna}) and for
proving some general statements, such as deriving the
Novikov-Shifman-Vainshrein-Zakharov (NSVZ) beta-function
\cite{Novikov:1983uc,Jones:1983ip,Novikov:1985rd,Shifman:1986zi} and
NSVZ-like relations in all orders of the perturbation theory
\cite{Stepanyantz:2011jy,Stepanyantz:2014ima,Shifman:2014cya,Shifman:2015doa}
or constructing in all orders the NSVZ-scheme
\cite{Kataev:2013eta,Kataev:2013csa,Kataev:2014gxa}. In particular,
it turns out that the higher derivative regularization has some
essential advantages comparing with the dimensional reduction.

${\cal N}=2$ supersymmetric theories can be certainly considered as
a special case of ${\cal N}=1$ supersymmetric theories with extra
hidden on-shell ${\cal N}=1$ supersymmetry. However, it is unclear
from the very beginning that the ${\cal N}=1$ higher covariant
derivative regularization will preserve the above hidden
supersymmetry. The first attempt to construct a version of the
higher derivative regularization for the ${\cal N}=2$ SYM theories
was made in \cite{Krivoshchekov:1985pq}, but the invariant higher
derivative term was not written explicitly. The problem was again
addressed in \cite{Buchbinder:2014wra}, where the higher covariant
derivative regularization was constructed for an arbitrary ${\cal
N}=2$ supersymmetric gauge theory. However, the formulation in terms
of ${\cal N}=1$ superfields which was used in
\cite{Buchbinder:2014wra} although preserve manifest ${\cal N}=1$
supersymmetry, does not allow to preserve the  hidden ${\cal N}=1$
supersymmetry at all stages of quantum corrections calculating,
because the gauge fixing term and ghosts have only manifest ${\cal
N}=1$ supersymmetry. It looks like the gauge fixing condition in
terms of ${\cal N}=1$ superfields is incompatible with hidden
supersymmetry. As a result, a removal of the above hole in the proof
of the ${\cal N}=2$ non-renormalization theorem requires the
additional study. It is clear that the most natural way to carry out
such a study should be based on a formulation of the ${\cal N}=2$
supersymmetric theories in terms of unconstrained ${\cal N}=2$
superfields where ${\cal N}=2$ supersymmetry will be manifest.

It is known that the manifest ${\cal N}=2$ supersymmetric
formulation of the ${\cal N}=2$ theories is given in the terms of
the ${\cal N}=2$ harmonic superspace
\cite{Galperin:1984av,Galperin:1985bj,Galperin:1985va} (see also
\cite{Galperin:2001uw}). In particular, using this formalism it is
possible to construct the ${\cal N}=2$ supersymmetric gauge fixing
procedure. That is why in this paper we formulate the higher
covariant derivative regularization for ${\cal N}=2$ supersymmetric
gauge theories in ${\cal N}=2$ harmonic superspace. As a result, we
obtain a version of the higher covariant derivative regularization
which allows to calculate quantum corrections in a manifestly ${\cal
N}=2$ supersymmetric way. Existence of such a regularization
justifies the proof of the ${\cal N}=2$ non-renormalization theorem.
Therefore, we present a way of calculating the quantum corrections
which actually ensures absence of divergences beyond the one-loop
approximation.

The paper is organized as follows: In Sect.
\ref{Section_Harmonic_Superspace} we recall basic information about
the ${\cal N}=2$ supersymmetric gauge theories and ${\cal N}=2$
harmonic superspace. Sect. \ref{Section_Higher_Derivatives} is
devoted to the formulation of the higher covariant derivative
regularization in the harmonic superspace. This is done by using the
background field method so that the constructed regularization does
not break the background gauge invariance. This allows to justify
the proof of the non-renormalization theorem, which is considered in
Sect. \ref{Subsection_Index}. In Sect.
\ref{Section_Non-renormalization} we present another simple proof of
${\cal N}=2$ non-renormalization theorem based on the NSVZ
$\beta$-function. The last Sect. \ref{Section_1-Loop} is devoted to
explicit calculating the one-loop divergences for the general ${\cal
N}=2$ SYM theory with matter by the help of the regularization
constructed in this paper. In particular, we demonstrate
factorizations of integrals for the $\beta$-function into integrals
of double total derivatives and vanishing of the one-loop anomalous
dimensions for hypermultiplets.

\section{${\cal N}=2$ supersymmetric gauge theories in the harmonic superspace}
\hspace{\parindent} \label{Section_Harmonic_Superspace}

Manifest ${\cal N}=2$ supersymmetry at all stages of calculating
quantum corrections is achieved by using ${\cal N}=2$ harmonic
superspace. It is obtained from the ordinary ${\cal N}=2$ superspace
with the coordinates $(x^\mu,\theta_{a}^i, \bar\theta_{i\dot a})$
\footnote{In our notation, $a$ numerates components of the left
spinor, $\dot a$ numerates components of the right spinor, and the
index $i=1,2$ numerates $\theta$-s.} by adding the complex
coordinates $u^\pm_i$, $u^-_i = (u^{+i})^*$, such that

\begin{equation}\label{U_Sphere}
u^{+i} u^{-}_{i}=1.
\end{equation}

In the language of ${\cal N}=2$ harmonic superspace the gauge field is a component of the real (with respect to a specially defined conjugation\, $\widetilde{}$\,) analytic superfield $V^{++}$.  The analyticity means that it satisfies the conditions

\begin{equation}
D_{a}^+ V^{++} = 0;\qquad \bar D_{\dot a}^+ V^{++} = 0,
\end{equation}

\noindent where $D_a^+$ and $\bar D_{\dot a}^+$ are the supersymmetric covariant derivatives contracted
with $u_i^+$. The superfield $iV^{++}$ belongs to the Lie algebra of the gauge group so that $V^{++} = e_0 (V^{++})^A t^A$, where $e_0$ is a bare coupling constant and the Hermitian generators $t^A$ are normalized by the condition $\mbox{tr}(t^A t^B) = \delta^{AB}/2$.
In order to write the action for the ${\cal N}=2$ SYM theory we also define the superfield

\begin{equation}\label{V--}
V^{--}(X,u) = \sum\limits_{n=1}^\infty (-i)^{n+1} \int du_1 du_2 \ldots
du_n \frac{V^{++}(X,u_1) V^{++}(X,u_2) \ldots V^{++}(X,u_n)}{(u^+
u_1^+)(u_1^+ u_2^+) \ldots (u_n^+ u^+)},
\end{equation}

\noindent
where $X$ denotes the set of the coordinates
$(x^\mu,\theta^i,\bar\theta_i)$, which is the same in all $V^{++}$
in the numerator, and $(u^+_\alpha u^+_\beta)\equiv u^{+i}_\alpha
u^{+}_{\beta i}$. This superfield is related to the strength tensors ${\cal W}$ and $\overline{\cal W}$ by the equations\footnote{Throughout this paper we mostly work in the $\lambda$-frame and omit the subscript $\lambda$ for the superfields in the $\lambda$-frame. The subscript $\tau$ points out that a superfield is written in the $\tau$-frame.}

\begin{equation}\label{W_Vs_V}
{\cal W} \equiv e^{iv} {\cal W}_\tau e^{-iv} = -\frac{i}{2} (\bar
D^+)^2 V^{--};\qquad \overline{\cal W} \equiv e^{iv}
\overline{\cal W}_\tau e^{-iv} = \frac{i}{2} (D^+)^2 V^{--}.
\end{equation}

\noindent In our notation $(\bar D^+)^2 \equiv \bar D^{+\dot a} \bar D^+_{\dot a}$, $(D^+)^2 \equiv D^{+a} D^+_{a}$, and the bridge superfield $v$ is defined as a solution of the equation

\begin{equation}\label{V_Vs_Bridge}
V^{++} \equiv -i e^{iv} D^{++} e^{-iv},
\end{equation}

\noindent
where

\begin{equation}\label{Harmonic_Derivatives}
D^{++} = u^{+i} \frac{\partial}{\partial u^{-i}};\qquad D^{--} =
u^{-i} \frac{\partial}{\partial u^{+i}}.
\end{equation}

\noindent It is important that the superfields ${\cal W}_\tau$ and $\overline{\cal W}_\tau$ depend only on the coordinates of ordinary superspace and are independent of the harmonic variables, $D^{\pm\pm}{\cal W}_\tau = 0$.\footnote{In the $\lambda$-frame this equation can be written as $D^{\pm\pm}{\cal W} + i[V^{\pm\pm},{\cal W}] =0$.}

The action of the pure ${\cal N}=2$ SYM theory in ${\cal N}=2$ harmonic superspace has the form \cite{Zupnik:1986ca,Zupnik:1987vm}

\begin{eqnarray}\label{Action_SYM}
&&\hspace*{-3mm} S_{\mbox{\scriptsize SYM}} = - \frac{1}{32 e_0^2}
\mbox{Re}\,\mbox{tr}\int d^4x\,d^2\theta_1\,d^2\theta_2\,{\cal
W}_\tau^2 = - \frac{1}{32 e_0^2} \mbox{Re}\,\mbox{tr}\int
d^4x\,d^2\theta_1\,d^2\theta_2\,du\,{\cal W}^2\nonumber\\
&&\hspace*{-3mm} = \frac{1}{16e_0^2} \sum\limits_{n=2}^\infty
\frac{(-i)^n}{n} \mbox{tr} \int d^4x\, d^8\theta\, du_1 du_2
\ldots du_n \frac{V^{++}(X,u_1) V^{++}(X,u_2) \ldots
V^{++}(X,u_n)}{(u_1^+ u_2^+)(u_2^+ u_3^+) \ldots (u_n^+
u_1^+)}.\qquad
\end{eqnarray}

\noindent This action is invariant under the gauge
transformations

\begin{equation}\label{Gauge_Transformations_V}
V^{++}\to e^{-i\lambda} V^{++} e^{i\lambda} - i e^{-i\lambda}
D^{++} e^{i\lambda},
\end{equation}

\noindent where $\lambda$ is a real (with respect to\, $\widetilde{}$\,) analytic superfield. Under these transformations

\begin{equation}\label{Gauge_Transformations_W}
V^{--}\to e^{-i\lambda} V^{--} e^{i\lambda} - i e^{-i\lambda}
D^{--} e^{i\lambda};\qquad {\cal W} \to e^{-i\lambda} {\cal W} e^{i\lambda};\qquad \overline{\cal W} \to e^{-i\lambda} \overline{\cal W} e^{i\lambda}.
\end{equation}

The general renormalizable ${\cal N}=2$ supersymmetric gauge model
consists of the pure Yang--Mills theory and hypermultiplets in a
certain representation of the gauge group. In ${\cal N}=2$
harmonic superspace the hypermultiplets are described by the
analytic superfields $\phi^+$. The action for the hypermultiplet with the mass $m_0$ can be
written as

\begin{equation}\label{Action_Hypermultiplet}
S_{\mbox{\scriptsize matter}} = - \int d^4x\,d^4\theta^+\, du\,
\widetilde\phi^+ \Big(D^{++} + iV^{++} - m_0 (\theta^+)^2 + m_0
(\bar\theta^+)^2\Big) \phi^+.
\end{equation}

\noindent In spite of the manifest dependence on $\theta^+$ and
$\bar\theta^+$ this action is ${\cal N}=2$ supersymmetric, because
for the massive representation corresponding to the hypermultiplet
supersymmetry algebra is modified by the central charge $Z=m_0$ (see,
e.g., \cite{Buchbinder:2001wy}).

The action (\ref{Action_Hypermultiplet}) is invariant under the
gauge transformations (\ref{Gauge_Transformations_V}) complemented
by the transformation of the hypermultiplet superfield

\begin{equation}
\phi^+ \to e^{-i\lambda} \phi^+; \qquad \widetilde\phi^+ \to
\widetilde\phi^+ e^{i\lambda}.
\end{equation}

\section{${\cal N}=2$ higher covariant derivative regularization}
\label{Section_Higher_Derivatives}

\subsection{${\cal N}=2$ higher derivative term}
\hspace{\parindent}

Let us considered the general ${\cal N}=2$ supersymmetric theory
described by the action

\begin{equation}
S = S_{\mbox{\scriptsize SYM}} + S_{\mbox{\scriptsize matter}},
\end{equation}

\noindent where $S_{\mbox{\scriptsize SYM}}$ is given by Eq.
(\ref{Action_SYM}) and $S_{\mbox{\scriptsize matter}}$ is given by
Eq. (\ref{Action_Hypermultiplet}), assuming that the analytical
superfield $\phi^+$ lies in an arbitrary representation $R$ of the
gauge group. In order to introduce the higher covariant derivative
regularization we add to the action the ${\cal N}=2$
supersymmetric higher derivative term

\begin{equation}\label{Higher_Derivative_Term}
S_{\Lambda} = - \frac{1}{128 e_0^2\Lambda^2}\mbox{tr} \int
d^4x\,d^8\theta\, \overline{\cal W}_\tau {\cal W}_\tau = - \frac{1}{128 e_0^2\Lambda^2}\mbox{tr} \int
d^4x\,d^8\theta\,du\, \overline{\cal W} {\cal W},
\end{equation}

\noindent
which is evidently also invariant under the gauge transformations (\ref{Gauge_Transformations_V}).
One can show that the expression
(\ref{Higher_Derivative_Term}) (up to notation) coincides with the
higher derivative term which was obtained in
\cite{Buchbinder:2014wra} by using the Noether method for ${\cal
N}=1$ superfields.\footnote{${\cal N}=1$ superfields are defined as
lowest components of ${\cal W}_\tau$ by the equations
\cite{deWit:1996kc,Banin:2002mf} ${\cal W}_\tau|\equiv 2\sqrt{2}\,
e^\Omega \Phi e^{-\Omega}$ and $(\nabla_2)^a {\cal W}_\tau|\equiv -
4 e^{\Omega} W^a e^{-\Omega}$, where the vertical line denotes the
conditions $\theta^2 = 0$ and $\bar\theta_2 = 0$.}

\subsection{The background field method and the gauge fixing procedure}
\hspace{\parindent}

In the case of using ${\cal N}=2$ harmonic superspace one can fix a gauge without breaking manifest ${\cal N}=2$
supersymmetry. It is convenient to do this using the background
field method. In the harmonic superspace it can be formulated as follows
\cite{Buchbinder:1997ya,Buchbinder:2001wy}. First, we split the analytic gauge superfield $V^{++}$ into the background and quantum parts
by making the substitution

\begin{equation}\label{Background_Splitting}
V^{++} = \bm{V}^{++} + v^{++}.
\end{equation}

\noindent
Then we can fix the gauge without breaking the background gauge invariance

\begin{equation}\label{Background_Invariance}
\bm{V}^{++} \to e^{-i\lambda} \bm{V}^{++} e^{i\lambda} - i
e^{-i\lambda} D^{++} e^{i\lambda};\qquad v^{++} \to e^{-i\lambda}
v^{++} e^{i\lambda};\qquad \phi^+ \to e^{-i\lambda} \phi^+
\end{equation}

\noindent
by inserting into the generating functional

\begin{equation}
1 = \Delta_{\mbox{\scriptsize FP}} \delta\left( \bm{\nabla}^{++}
v^{++} - f^{(+4)}\right),
\end{equation}

\noindent where the background covariant derivative is given by

\begin{equation}
\bm{\nabla}^{++} v^{++} \equiv D^{++} v^{++} + i
[\bm{V}^{++},v^{++}].
\end{equation}

\noindent It is well known \cite{Buchbinder:2001wy} that in this
case the Faddeev--Popov determinant $\Delta_{\mbox{\scriptsize
FP}}$ can be presented as a functional integral over the
Faddeev--Popov ghosts, which are described by the anticommuting
analytical superfields $b$ (antighost) and $c$ (ghost):

\begin{equation}
\Delta_{\mbox{\scriptsize FP}} = \int Db Dc
\exp(iS_{\mbox{\scriptsize FP}}),
\end{equation}

\noindent
where the action for the Faddeev--Popov ghosts is given by

\begin{equation}
S_{\mbox{\scriptsize FP}} = \frac{1}{e_0^2} \mbox{tr}\int
d^4x\,d^4\theta^+ du\, b\, \bm{\nabla}^{++} \Big(\bm{\nabla}^{++}
c + i[v^{++},c]\Big).
\end{equation}

\noindent
(The ghost superfields $b$ and $c$ belong to the adjoint representation of the gauge group.)
Then it is convenient to integrate over $f^{+4}$ taking into account the identity

\begin{eqnarray}\label{Nielsen_Kallosh_Determinant}
&& 1 = \Delta_{\mbox{\scriptsize NK}} \int Df^{(+4)}
\exp\Big(-\frac{i}{32\xi_0 e_0^2}\mbox{tr}
\int d^4x\,d^8\theta\, du_1 du_2\nonumber\\
&&\qquad\qquad\qquad\qquad\qquad\qquad \times
e^{-i\mbox{\boldmath$v$}_1} f^{(+4)}_1 e^{i\mbox{\boldmath$v$}_1}
\frac{(u_1^- u_2^-)}{(u_1^+ u_2^+)^3}\,
e^{-i\mbox{\boldmath$v$}_2} \Big(1+
\frac{\stackrel{\bm{\smile}}{\bm{\Box}}_2}{\Lambda^2}\Big)
f^{(+4)}_2 e^{i\mbox{\boldmath$v$}_2}\Big).\qquad
\end{eqnarray}

\noindent Here the subscripts numerates the harmonic variables,
e.g., $f^{(+4)}_1 \equiv f^{(+4)}(X,u_1)$ etc. The superfield

\begin{equation}
\bm{v} = v\Big|_{v^{++}=0}
\end{equation}

\noindent is introduced in order to obtain the expression
invariant under the background gauge transformations
(\ref{Background_Invariance}), under which

\begin{equation}\label{Background_Invariance_Bridge}
e^{i\mbox{\boldmath$v$}} \to e^{-i\lambda} e^{i\mbox{\boldmath$v$}} e^{i\tau},
\end{equation}

\noindent where $\tau = \tau(x,\theta)$ is independent of the
harmonic variables. The bridge superfield $\bm{v}$ is related with the background
gauge superfields by the equations

\begin{equation}
\bm{V}^{++} = -i e^{i\mbox{\boldmath$v$}} D^{++}
e^{-i\mbox{\boldmath$v$}};\qquad\qquad \bm{V}^{--} = -i
e^{i\mbox{\boldmath$v$}} D^{--} e^{-i\mbox{\boldmath$v$}}.
\end{equation}

Also in Eq. (\ref{Nielsen_Kallosh_Determinant}) we use the
notation

\begin{equation}
\stackrel{\bm{\smile}}{\bm{\Box}}\, \equiv -\frac{1}{32} (D^+)^4
(\bm{\nabla}^{--})^2
\end{equation}

\noindent for the analog of the Laplace operator, which maps
analytic superfields into analytic superfields, where the
background covariant derivative is given by $\bm{\nabla}^{--}
\equiv D^{--} + i \bm{V}^{--}$.

Inserting the expression (\ref{Nielsen_Kallosh_Determinant}) into
the generating functional corresponds to adding the gauge fixing
action

\begin{equation}\label{Gauge_Fixing_Term}
S_{\mbox{\scriptsize gf}} = -\frac{1}{32\xi_0 e_0^2}\mbox{tr} \int
d^4x\,d^8\theta\, du_1 du_2\, e^{-i\mbox{\boldmath$v$}_1}
\bm{\nabla}^{++}_1 v^{++}_1 e^{i\mbox{\boldmath$v$}_1}
\frac{(u_1^- u_2^-)}{(u_1^+ u_2^+)^3}\,
e^{-i\mbox{\boldmath$v$}_2} \Big(1+
\frac{\stackrel{\bm{\smile}}{\bm{\Box}}_2}{\Lambda^2}\Big)
\bm{\nabla}_2^{++} v^{++}_2 e^{i\mbox{\boldmath$v$}_2},
\end{equation}

\noindent which is invariant under
the background gauge transformations (\ref{Background_Invariance}) and (\ref{Background_Invariance_Bridge}).

Using the equation

\begin{equation}\label{Laplace}
-\frac{1}{32} (D^+)^4 (D^{--})^2 v^{++} = \partial^2 v^{++}
\end{equation}

\noindent
one can verify that the terms quadratic in the quantum superfield $v^{++}$ (which do not contain the background
superfield) can be written as

\begin{eqnarray}
&& S^{(2)}_{\mbox{\scriptsize SYM}} + S^{(2)}_\Lambda +
S^{(2)}_{\mbox{\scriptsize gf}} = - \frac{1}{8 e_0^2\xi_0}\mbox{tr}
\int d^4x\,d^4\theta^+\,du\, v^{++}(X,u)\,
\partial^2 \Big(1+\frac{\partial^2}{\Lambda^2}\Big) v^{++}(X,u)\qquad\nonumber\\
&& + \frac{1}{32 e_0^2}\Big(1-\frac{1}{\xi_0}\Big)\mbox{tr} \int d^4x\,d^8\theta\,du_1\, du_2\,
v^{++}(X,u_1) \frac{1}{(u_1^+ u_2^+)^2}
\Big(1+\frac{\partial^2}{\Lambda^2}\Big)\, v^{++}(X,u_2).
\end{eqnarray}

\noindent
In the case $\xi_0=1$ these terms have the most simple form

\begin{equation}
- \frac{1}{8e_0^2}\mbox{tr} \int
d^4x\,d^4\theta^+\,du\,v^{++}
\partial^2 \Big(1+\frac{\partial^2}{\Lambda^2}\Big) v^{++}.
\end{equation}

Following Ref. \cite{Buchbinder:2001wy}, one can easily calculate
the Nielsen--Kallosh determinant $\Delta_{\mbox{\scriptsize NK}}$.
It is given by a product of two contributions, one of which can be
presented as an integral over the commuting analytic
Nielsen--Kallosh superfield $\beta$ in the adjoint representation
of the gauge group

\begin{equation}
\Delta_{\mbox{\scriptsize NK}} = \int D\beta
\exp(iS_{\mbox{\scriptsize NK}}) \cdot \mbox{Det}^{1/2}(NK;
\bm{V}^{++}),
\end{equation}

\noindent where

\begin{equation}
S_{\mbox{\scriptsize NK}} = \frac{1}{e_0^2} \mbox{tr}\int
d^4x\,d^4\theta^+ du\, \bm{\nabla}^{++} \beta \bm{\nabla}^{++}
\beta.
\end{equation}

\noindent The second determinant can be also presented as a
functional integral over anticommuting analytic superfields in the
adjoint representation $\gamma^{(+4)}$ and $\gamma$:

\begin{equation}\label{Second_Determinant}
\mbox{Det}(NK; \bm{V}^{++}) = \int D\gamma^{(+4)} D\gamma\,
\exp\Big\{\frac{i}{e_0^2} \mbox{tr}\int d^4x\,d^4\theta^+ du\,
\gamma^{(+4)} \stackrel{\bm{\smile}}{\bm{\Box}} \Big(1+
\frac{\stackrel{\bm{\smile}}{\bm{\Box}}}{\Lambda^2}\Big)
\gamma\Big\},
\end{equation}

\noindent but the degree $1/2$ does not allow to modify
$S_{\mbox{\scriptsize NK}}$ in such a way to include this
contribution.

\subsection{Degree of divergence and the non-renormalization theorem}
\hspace{\parindent}\label{Subsection_Index}

In this subsection we will evaluate the superficial degree of divergence for an arbitrary global ${\cal N}=2$ supersymmetric
gauge theory and prove that any such theory is finite beyond
the one-loop approximation. The analysis is based on two properties.
First, the effective action is manifestly ${\cal N}=2$
supersymmetric. It is stipulated by manifest ${\cal N}=2$
supersymmetry of the theory regularized by the ${\cal
N}=2$ supersymmetric higher covariant derivative
regularization\footnote{Namely, the manifest ${\cal N}=2$
supersymmetry was assumed but not proved in all other regularization
schemes.}. Second, the regularized effective action is manifestly
gauge invariant. It is stipulated by background field method
developed in the previous subsection. Therefore, for evaluating the
superficial degree of divergence $\omega$ we can use manifest
${\cal N}=2$ supersymmetry and the manifest gauge invariance. Also
we will take into account the discussion of the degree of
divergence carried out in \cite{Buchbinder:1997ib}.

Let us study an arbitrary $L$-loop supergraph ($L>1$) and set $m_0=0$, because masses cannot
increase the degree of divergence. In the beginning, we will consider the limit
$\Lambda\to \infty$ which corresponds to the non-regularized theory. In this case the momentum integrals do
not contain any dimensionful parameters and the degree of divergence
can be calculated using dimensional considerations. Calculating the
contribution to the effective action of a certain supergraph we
obtain the integral over $d^8\theta$ and all external momenta. It is
easy to see that in the coordinate representation the dimensions of
the gauge superfield, the hypermultiplet, and the Faddeeev--Popov
ghosts are $[V^{++}(x,\theta,u)]=m^{0}$, $[\phi(x,\theta,u)]=m^{1}$,
and $[b(x,\theta,u)]=[c(x,\theta,u)]=m^{1}$. Therefore, in the
momentum representation $[V(p,\theta,u)]=m^{-4}$ and
$[\phi(p,\theta,u)]=[b(p,\theta,u)]=[c(p,\theta,u)]=m^{-3}$. As a
consequence, the dimension of the integral over $d^8\theta$ and
external lines (including the corresponding momentum integrals) is
$m^{(4+N_\phi+N_c)}$, where $N_\phi$ and $N_c$ are numbers of the
hypermultiplet and ghost external legs, respectively. The dimension
of the momentum $\delta$-function (which leads to the
energy--momentum conservation) is $m^{-4}$. Moreover, if there are
$N_D$ spinor derivatives acting to the external gauge lines, they
give a factor of the dimension $m^{N_D/2}$. Taking into account that
effective action is dimensionless, we obtain that the dimension of
the remaining momentum integral (which is equal to the degree of
divergence for the non-regularized theory) is
\cite{Buchbinder:1997ib}

\begin{equation}\label{Index}
\omega = - N_\phi - N_c - \frac{1}{2} N_D.
\end{equation}

Now, let us proceed to calculating the degree of divergence for the theory containing the higher derivative term (\ref{Higher_Derivative_Term}).
Due to the presence of this term the degree of momentums in the
denominator of the gauge propagator is increased by 2. Also the degree
of momentums in the purely gauge vertices is increased by 2. Therefore, in the
regularized theory the degree of divergence is given by

\begin{equation}\label{Regularized_Index}
\omega_\Lambda = - N_\phi - N_c - \frac{1}{2} N_D - 2(P-V),
\end{equation}

\noindent where $V$ is a number of the purely gauge vertices and $P$
is a number of the gauge propagators. If the regularized effective
action is manifestly ${\cal N}=2$ supersymmetric\footnote{Existence
of this property was assumed in \cite{Buchbinder:1997ib}, however
the regularization scheme which provided such a property was not
proposed. Here we eliminate this hole in proof of the ${\cal N}=2$
non-renormalization theorem.} and formulated on the base of
background field method, the quantity $N_D$, associated with
external vector superfield lines, is always positive beyond the
one-loop (see discussion of this point in \cite{Buchbinder:1997ib}).
Evidently, beyond the one-loop approximation we have $P-V>0$ and,
therefore, in this case $\omega_\Lambda < 0$.

In one-loop approximation ($L=1$) the effective action is given by
the functional determinants of the differential operators acting on
superfields and requires a separate consideration. If the background
field is included into the propagator as in Ref.
\cite{Buchbinder:1997ib}, then the one-loop diagrams do not contain
external lines and $N_D=0$. Therefore, $\omega=0$. This implies that
the divergencies in principle can be present in the one-loop
diagrams (including the one-loop subdiagrams in multiloop diagrams).
As a consequence, one-loop divergencies cannot be removed by adding
the regularizing higher derivative term that is a typical feature of
the higher covariant derivative regularization \cite{Faddeev:1980be}. In order to regularize them
by a manifestly ${\cal N}=2$ supersymmetric and gauge invariant way
one should introduce into the generating functional the appropriate
manifestly ${\cal N}=2$ supersymmetric and gauge invariant Pauli--Villars
determinants, as it was first done in \cite{Slavnov:1977zf} for conventional field theory.

\subsection{Removing one-loop divergences by the Pauli--Villars determinants}
\hspace{\parindent}

In this section we develop the harmonic superspace
Pauli-Villars regularization for the one-loop divergences which remain after adding the higher derivative term (\ref{Higher_Derivative_Term}) to the classical action.

In ${\cal N}=2$ harmonic superspace the Pauli--Villars determinants
are constructed using the expression for the action of the massive
hypermultiplet. Following Ref. \cite{Buchbinder:2014wra}, for this
purpose we introduce the (commuting) analytic Pauli--Villars
superfields $\varphi^+$ (in the adjoint representation of the gauge
group) and $\phi_I^+$ (which lies in the same representation as the
superfield $\phi^+$) and construct the Pauli--Villars determinants

\begin{eqnarray}\label{PV_Determinants}
&& \mbox{Det}(PV, M_0; V^{++})^{-1} = \int
D\varphi^+\,D\widetilde\varphi^+
\exp(i S_{\varphi});
\nonumber\\
&& \mbox{Det}(PV, M_I; V^{++})^{-1} = \int
D\phi_I^+\,D\widetilde\phi_I^+ \exp(i S_{I}),
\end{eqnarray}

\noindent where the actions for the Pauli--Villars fields are now
written as

\begin{eqnarray}
&& S_\varphi = - \frac{2}{e_0^2} \mbox{tr} \int
d^4x\,d^4\theta^+\, du\, \widetilde\varphi^+ \Big(D^{++} \varphi^+
+ i[V^{++}, \varphi^+] - M_0 (\theta^+)^2 \varphi^+ + M_0
(\bar\theta^+)^2 \varphi^+\Big);\qquad\nonumber\\
&& S_I = - \int d^4x\,d^4\theta^+\, du\, \widetilde\phi_I^+
\Big(D^{++} + iV^{++} - M_I (\theta^+)^2 + M_I
(\bar\theta^+)^2\Big) \phi_I^+.
\end{eqnarray}

\noindent (It is assumed that the superfield $V^{++}$ is split
into the background and quantum parts acoording to Eq.
(\ref{Background_Splitting}).) The masses of the Pauli--Villars
superfields $M_0$ and $M_I$ are proportional to the parameter
$\Lambda$ in the higher derivative term, the coefficient of the
proportionality being independent of the (bare) coupling constant.

In the next section we demonstrate by explicit calculation that inserting the Pauli--Villars
determinants (\ref{PV_Determinants}) leads to regularizing all one-loop divergencies.

Using the Pauli--Villars determinants (\ref{PV_Determinants}) it is possible to construct the
regularized generating functional as

\begin{eqnarray}
&& Z = \int Dv^{++} D\widetilde\phi^+ D\phi^+\,Db\,Dc\, D\beta\,
\mbox{Det}(PV, M_0; V^{++})^{-1}\prod\limits_{I=1}^n
\mbox{Det}(PV, M_I; V^{++})^{c_I}\qquad\nonumber\\
&& \times \mbox{Det}^{1/2}(NK; \bm{V}^{++}) \exp\Big(iS +
iS_\Lambda  + i S_{\mbox{\scriptsize gf}} + i S_{\mbox{\scriptsize
ghosts}}+ i S_{\mbox{\scriptsize sources}}\Big),\qquad\quad
\end{eqnarray}

\noindent where $c_I$ are the coefficients which satisfy the
conditions $\sum_{I=1}^n c_I = 1$ and $\sum_{I=1}^n c_I M_I^2 = 0$.
The action $S$ is a sum of Eqs. (\ref{Action_SYM}) and
(\ref{Action_Hypermultiplet}), $S_\Lambda$ is the higher derivative
term (\ref{Higher_Derivative_Term}), $S_{\mbox{\scriptsize gf}}$ is
the gauge fixing term (\ref{Gauge_Fixing_Term}), and
$S_{\mbox{\scriptsize ghosts}} = S_{\mbox{\scriptsize FP}} +
S_{\mbox{\scriptsize NK}}$. The source term $S_{\mbox{\scriptsize
sources}}$ includes all necessary sources. The effective action is
defined by the standard way on the base of $Z$.

Thus, we obtain the ${\cal N}=2$ supersymmetric regularization which has never been
considered before and hope that it will be useful for various
concrete calculations.

\section{The exact NSVZ $\beta$-function and ${\cal N}=2$ non-renormalization theorem}
\hspace{\parindent}\label{Section_Non-renormalization}

The higher derivative regularization constructed in this paper
allows to reformulate a statement of the non-renormalization theorem
in terms of the NSVZ $\beta$-function
\cite{Shifman:1999mv,Buchbinder:2014wra}. The matter is that there
are strong evidences that the NSVZ relation is satisfied by the
renormalization group functions defined in terms of the bare
coupling constant if the higher covariant derivatives are used for
the regularization
\cite{Pimenov:2009hv,Stepanyantz:2011jy,Stepanyantz:2014ima,Shifman:2014cya,Shifman:2015doa}.
The ${\cal N}=2$ supersymmetric theories can be considered as a
special case of ${\cal N}=1$ supersymmetric theories. In
particular, for ${\cal N}=2$ gauge theories the NSVZ
$\beta$-function gives (see, e.g., \cite{Buchbinder:2014wra}).

\begin{equation}\label{Exact_Beta}
\beta(\alpha_0) = - \frac{\alpha_0^2}{\pi}\Big(C_2 -
T(R_0)\Big)\Big(1 - \gamma_\phi(\alpha_0)\Big),
\end{equation}

\noindent where $\gamma_\phi(\alpha_0)$ is the anomalous dimension
of the hypermultiplet. If the theory is formulated in terms of
${\cal N}=1$ superfields it is at least very difficult (if possible)
to prove that $\gamma_\phi=0$. However, this can be easily done
using the regularization constructed in this paper. Really, the
diagrams contributing to the anomalous dimension of the
hypermultiplet $\gamma_\phi(\alpha_0)$ have $N_\phi=2$ (see
Eq. (\ref{Index})), and, therefore, are finite. Thus, the
anomalous dimension vanishes and the $\beta$-function is given by
the purely one-loop expression

\begin{equation}\label{Exact_Beta_One-Loop}
\beta(\alpha_0) = - \frac{\alpha_0^2}{\pi}\Big(C_2 - T(R_0)\Big).
\end{equation}

\noindent Eq. (\ref{Exact_Beta_One-Loop}) and vanishing of the
anomalous dimension $\gamma_\phi$ imply that ${\cal N}=2$
supersymmetric gauge theories are finite beyond the one-loop
approximation, and the hypermultiplets are not renormalized. For the
${\cal N}=4$ SYM theory $R_0 = Adj$ and $T(R_0) = T(Adj) = C_2$. As
a consequence, we obtain the known results that the $\beta$-function
vanishes and the theory is finite in all orders.

\section{One-loop quantum corrections}
\hspace{\parindent} \label{Section_1-Loop}

According to the non-renormalization theorem considered in the
previous section, the divergences can appear only in the one-loop
approximation. Due to the background gauge invariance and
renormalizability these divergences are encoded in the
renormalization constants, so that the counterterms $\Delta S \equiv
S - S_{\mbox{\scriptsize ren}}$ can be presented in the form

\begin{eqnarray}\label{Counterterms}
&& \Delta S = -\frac{1}{32 e^2} \mbox{Re}\,\mbox{tr}\int
d^4x\,d^2\theta_1\,d^2\theta_2\,du\,\Big(Z_\alpha{\cal W}^2[\bm{V}^{++} + Z_v v_R^{++}] - {\cal W}^2[\bm{V}^{++} + v_R^{++}]\Big)\nonumber\\
&& + \frac{1}{e^2} \mbox{tr}\int
d^4x\,d^4\theta^+ du\, \Big((Z_c Z_\alpha-1) b_R\, \bm{\nabla}^{++} \bm{\nabla}^{++}
c_R  + i(Z_c Z_\alpha Z_v-1)b_R\, [v_R^{++},c_R]\Big) \nonumber\\
&& - \int d^4x\,d^4\theta^+\, du\,
\Big((Z_\phi-1) \widetilde\phi_R^+ \bm{\nabla}^{++} \phi_R^+ + i (Z_\phi Z_v -1) \widetilde\phi_R^+ v_R^{++} \phi_R^+\Big),\qquad
\end{eqnarray}

\noindent
where the subscript $R$ denotes the renormalized fields and $e$ is the renormalized coupling constant. By definition, the sum of $\Delta S$ and the divergent part of the effective action $\Gamma_\infty$ is finite. Thus, the renormalization constants $Z_\alpha$, $Z_\phi$, $Z_v$, and $Z_c$ completely define the divergent part of the effective action. In order to find these renormalization constants we can consider only
two-point Green functions of the various superfields using the above constructed version of the higher covariant derivative regularization in the harmonic superspace. (For simplicity, here we will consider the massless case $m_0=0$ and the gauge $\xi=1$.)

\begin{figure}[h]

\begin{picture}(0,1.5)
\put(4.0,0){\includegraphics[scale=0.4]{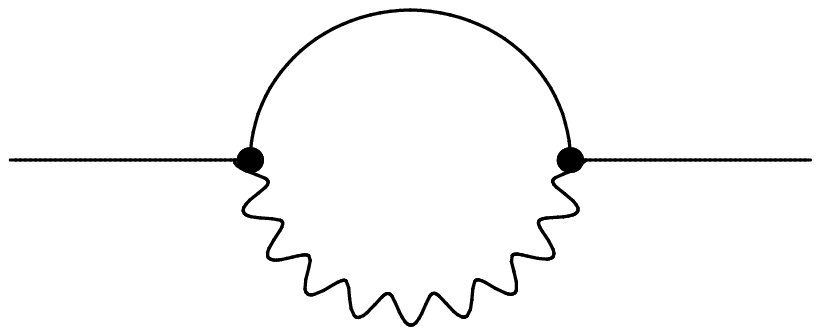}}
\put(9.0,0){\includegraphics[scale=0.4]{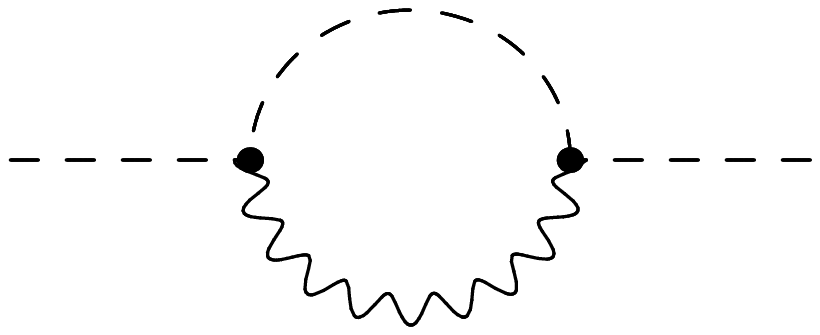}}
\end{picture}

\caption{One-loop diagrams which contribute to the two-point Green functions of the matter superfields and the Faddeev--Popov ghosts.}
\label{Figure_One_Loop_Gamma}
\end{figure}

First, we consider the two-point Green functions of the matter superfields and the Faddeev--Popov ghosts (which are given by the diagrams presented in Fig.
\ref{Figure_One_Loop_Gamma}). We obtained that these diagrams give the vanishing contributions similar to the calculation made in \cite{Galperin:1985va}. The only difference is the presence of higher derivatives in the propagator of the quantum gauge superfield. For example, the one-loop contribution to the two-point function of the hypermultiplet superfields is proportional to

\begin{equation}
\int \frac{d^4p}{(2\pi)^4} d^8\theta\, du\, \widetilde\phi^{+i}(p,\theta,u) C(R)_i{}^j D^{--} \phi_j^+(-p,\theta,u) \int \frac{d^4k}{(2\pi)^4} \frac{e_0^2}{k^4 (1+k^2/\Lambda^2) (k+p)^2} = 0.
\end{equation}

\noindent
(In order to derive the last equality we note that the integration measure contains $(D^{+})^4$ and $(D^+)^4 D^{--} \phi^+ = 0$ due to analyticity of $\phi^+$.)

As a consequence, $Z_\phi=1+O(\alpha_0^2)$ and $Z_c Z_\alpha=1+O(\alpha_0^2)$. This implies that in the considered approximation the anomalous dimension of the hypermultiplet vanishes,
$\gamma_\phi(\alpha_0) = O(\alpha_0^2)$.

\begin{figure}[h]

\begin{picture}(0,4)
\put(0.5,2.4){\includegraphics[scale=0.4]{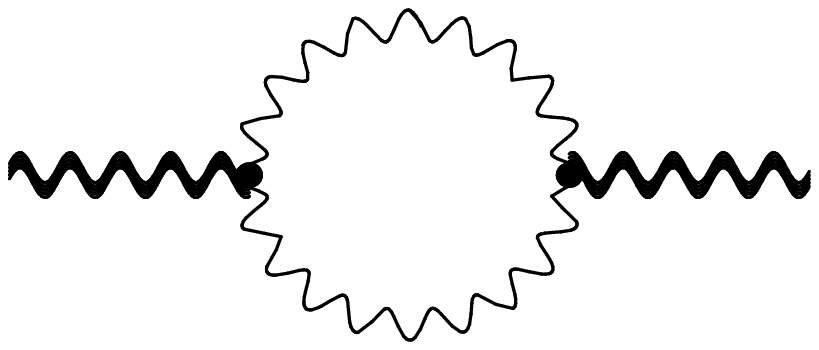}}
\put(0.9,0){\includegraphics[scale=0.4]{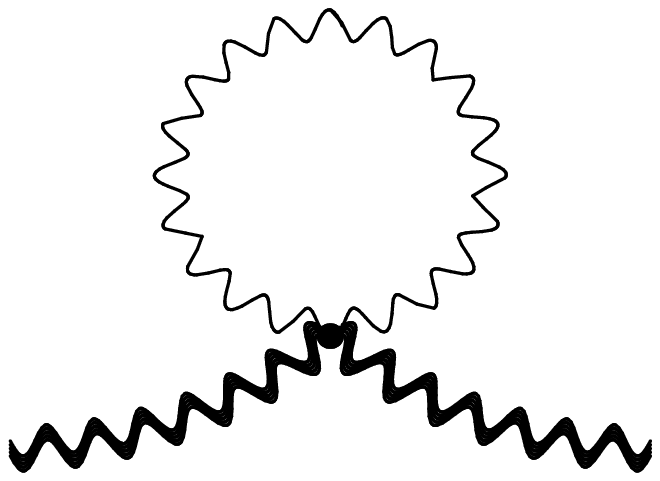}}
\put(4.3,2.45){\includegraphics[scale=0.4]{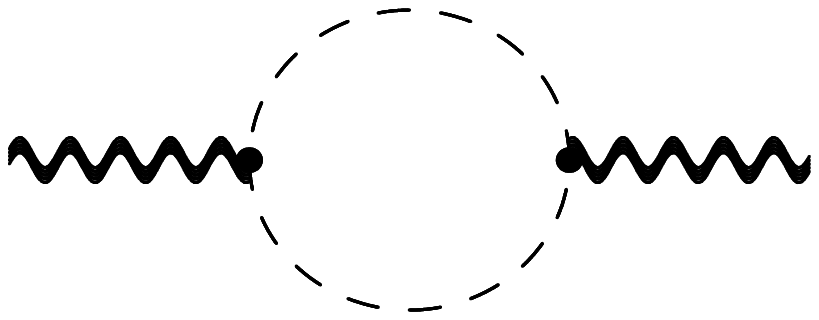}}
\put(4.6,0){\includegraphics[scale=0.4]{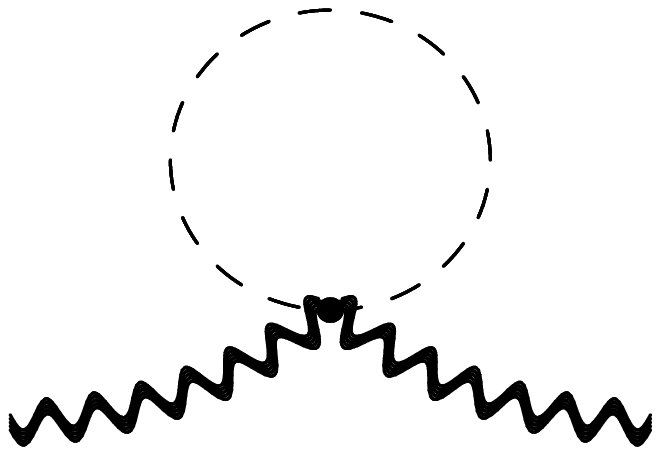}}
\put(8.1,2.4){\includegraphics[scale=0.4]{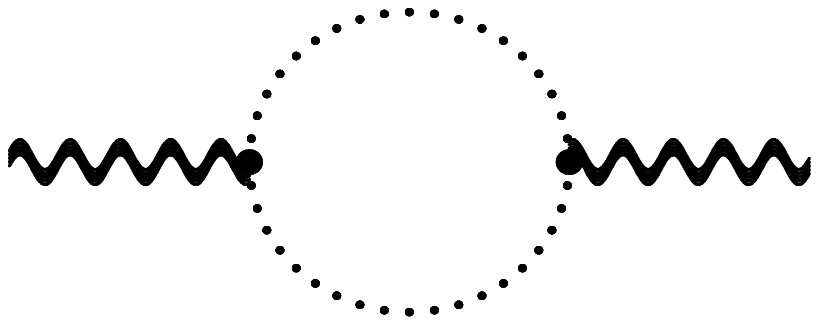}}
\put(8.5,0){\includegraphics[scale=0.4]{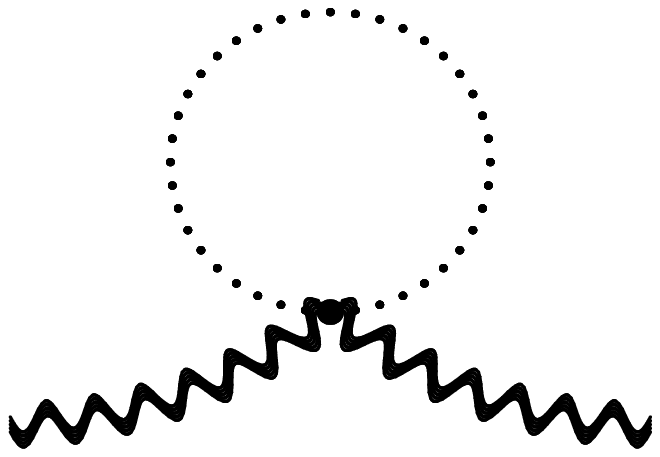}}
\put(12.0,2.43){\includegraphics[scale=0.4]{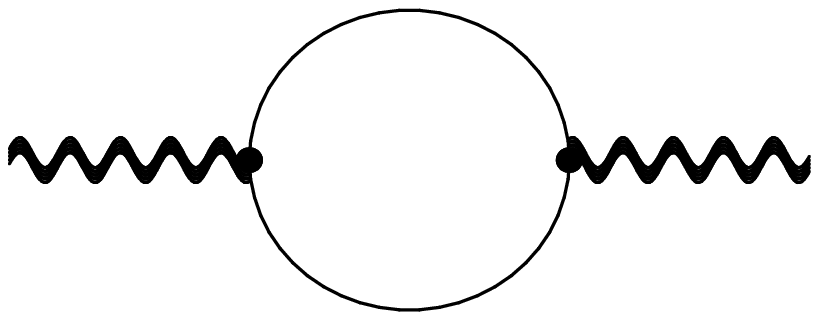}}
\end{picture}
\vspace*{1mm} \caption{One-loop diagrams which give the two-point
Green function of the background superfield. The external lines
correspond to the background gauge superfield $\bm{V}^{++}$.}
\label{Figure_One_Loop_Beta}
\end{figure}

Next, we consider the diagrams which give the one-loop renormalization of the coupling constant $Z_\alpha$. This renormalization constant can be found by calculating the two-point Green function of the background superfield $\bm{V}^{++}$ (which corresponds to the bold wavy external lines). The corresponding one-loop diagrams are presented in Fig. \ref{Figure_One_Loop_Beta}. The result can be written as

\begin{eqnarray}\label{Two_Point_V}
&& \frac{d\Gamma^{(2)}_{\bm{V}}}{d\ln\Lambda}\Big|_{\Lambda\to \infty} = \frac{1}{128\pi}\mbox{tr} \int d^8\theta\, du_1 du_2
\frac{1}{(u_1^+ u_2^+)^2} \int \frac{d^4p}{(2\pi)^4}
\bm{V}^{++}(-p,\theta,u_1) \bm{V}^{++}(p,\theta,u_2)\nonumber\\
&& \times \Big(I_{\mbox{\scriptsize gauge}}
+ I_{\mbox{\scriptsize FP}} + I_{\mbox{\scriptsize NK}} +
I_\varphi + I_\phi + O(\alpha_0)\Big),
\end{eqnarray}

\noindent where the derivative with respect to $\ln\Lambda$ is
calculated at a fixed value of the renormalized coupling
constant $\alpha$.

$I_{\mbox{\scriptsize gauge}}$ denotes the
contribution of the diagrams containing a loop of the quantum
gauge superfield presented in the first column of Fig.
\ref{Figure_One_Loop_Beta}. We have obtained

\begin{equation}
I_{\mbox{\scriptsize gauge}} = 0.
\end{equation}

\noindent
(In order to obtain this result it is necessary to take into account vertices containing higher derivatives which (in the one-loop approximation) cancel higher derivatives in the propagators. Thus, although the result is same as in the case in which the higher derivatives are absent, its derivation is essentially different.)

The second and the third columns in Fig.
\ref{Figure_One_Loop_Beta} contain diagrams with a loop of the
Faddeev--Popov ghosts and Nielsen--Kallosh ghosts, respectively.
Because the Faddeev--Popov ghosts are anticommuting, while the
Nielsen--Kallosh ghosts commute, we obtain $I_{\mbox{\scriptsize
FP}} = - 2 I_{\mbox{\scriptsize NK}}$, where we also take into
account that the determinant (\ref{Second_Determinant}) gave the
vanishing contribution. Both $I_{\mbox{\scriptsize FP}}$ and
$I_{\mbox{\scriptsize NK}}$ are not well-defined, but the
well-defined result is obtained after adding the loop of the
Pauli--Villars superfield $\varphi^+$. This contribution is given by
the diagram in the fourth column in Fig.
\ref{Figure_One_Loop_Beta}. Also this diagram gives a contribution
of the matter superfield $\phi^+$. After
calculating the diagrams in Fig. \ref{Figure_One_Loop_Beta} we
have obtained

\begin{eqnarray}
&& I_{\mbox{\scriptsize FP}} + I_{\mbox{\scriptsize NK}} +
I_\varphi = -8\pi C_2 \int \frac{d^4q}{(2\pi)^4}
\frac{d}{d\ln\Lambda}
\Big(\frac{1}{q^4} - \frac{1}{(q^2+M_0^2)^2}\Big)\nonumber\\
\nonumber\\
&&\qquad\qquad\qquad\ \  = 2\pi C_2\int \frac{d^4q}{(2\pi)^4} \frac{\partial}{\partial q^\mu} \frac{\partial}{\partial q_\mu}
\frac{d}{d\ln\Lambda} \Big[\frac{1}{q^2}\Big(\ln q^2 - \ln (q^2+M_0^2) \Big)\Big]
= -\frac{C_2}{\pi};\qquad\\
&& I_{\phi} = 8\pi T(R)\int \frac{d^4q}{(2\pi)^4}
\frac{d}{d\ln\Lambda} \Big(\frac{1}{q^4} - \sum\limits_{I=1}^n c_I
\frac{1}{(q^2+M_I^2)^2}\Big)
\nonumber\\
&&\qquad = -2\pi T(R)\int \frac{d^4q}{(2\pi)^4} \frac{\partial}{\partial q^\mu} \frac{\partial}{\partial q_\mu}
\frac{d}{d\ln\Lambda} \Big[\frac{1}{q^2}\Big(\ln q^2 - \sum\limits_{I=1}^n c_I \ln (q^2+M_I^2) \Big)\Big]
= \frac{T(R)}{\pi}.\qquad\quad
\end{eqnarray}

\noindent
(Calculating these integrals we take into account that the masses of the Pauli--Villars superfields
$M_0$ and $M_I$ are proportional to the parameter $\Lambda$.)
Thus, both these integrals are well-defined integrals of double total derivatives. (This is a typical feature obtained if supersymmetric theories are regularized by higher covariant derivatives, which was first noted in \cite{Soloshenko:2003nc,Smilga:2004zr}.)
Substituting the results for these integrals into Eq. (\ref{Two_Point_V}) we obtain

\begin{equation}
Z_\alpha = 1 + \frac{\alpha_0}{\pi}\Big(C_2 - T(R_0)\Big) \ln \frac{\Lambda}{\mu} + O(\alpha_0^2).
\end{equation}

\noindent
As a consequence, in the considered approximation

\begin{equation}
\frac{\beta(\alpha_0)}{\alpha_0^2} = -\frac{1}{\alpha_0} \frac{d\ln Z_\alpha}{d\ln\Lambda} =  - \frac{1}{\pi}\Big(C_2 - T(R_0)\Big) + O(\alpha_0).
\end{equation}

\begin{figure}[h]

\begin{picture}(0,2)
\put(0.6,0.2){\includegraphics[scale=0.4]{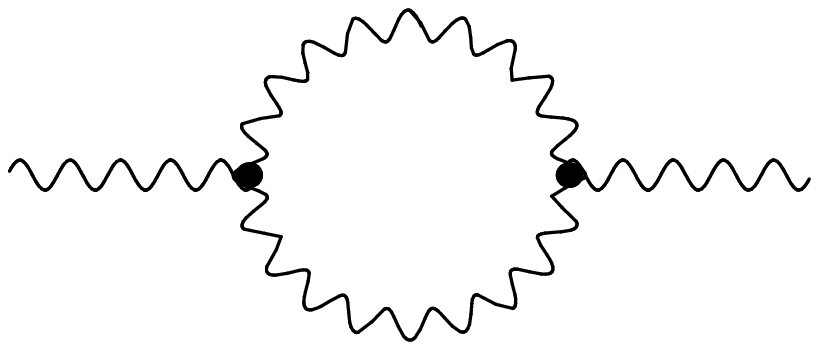}}
\put(4.6,0){\includegraphics[scale=0.4]{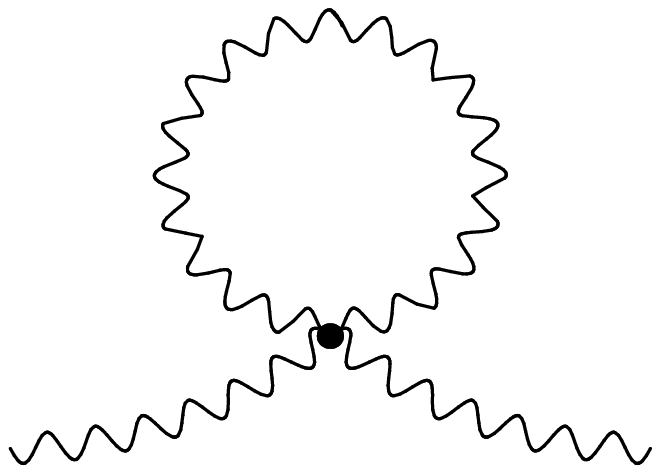}}
\put(7.8,0.25){\includegraphics[scale=0.4]{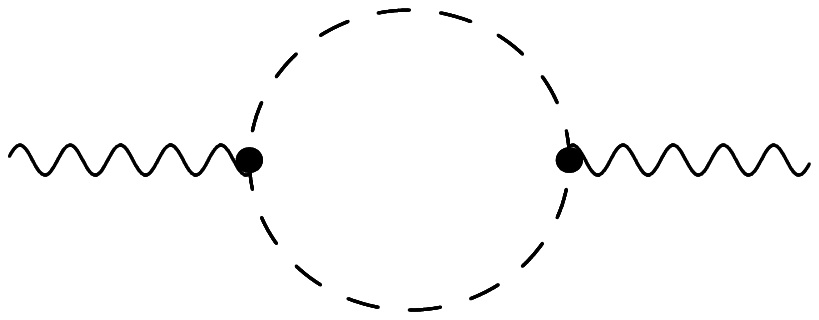}}
\put(11.8,0.25){\includegraphics[scale=0.4]{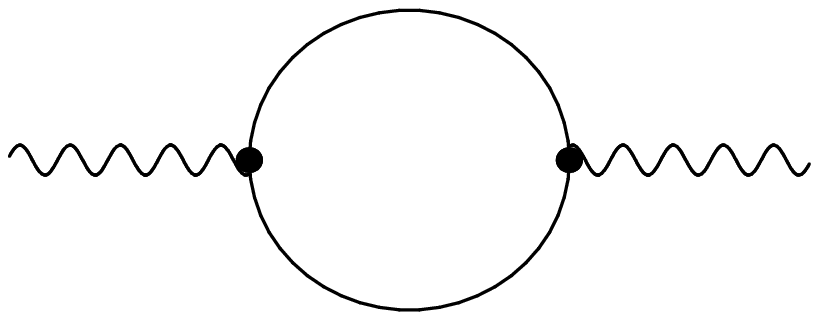}}
\end{picture}
\vspace*{1mm} \caption{One-loop diagrams which give the two-point
Green function of the quantum gauge superfield $v^{++}$.}
\label{Fiqure_Quantum_Superfield_Renormalization}
\end{figure}

The renormalization constant $Z_v$ in the one-loop approximation can be found by calculating the diagrams presented in Fig.
\ref{Fiqure_Quantum_Superfield_Renormalization}.

The corresponding contribution to the effective action has the form

\begin{eqnarray}\label{Quantum_Green_Function}
&& \frac{d\Gamma^{(2)}_{v}}{d\ln\Lambda}\Big|_{\Lambda\to \infty} = \frac{1}{128\pi}\mbox{tr} \int d^8\theta\, du_1 du_2
\frac{1}{(u_1^+ u_2^+)^2} \int \frac{d^4p}{(2\pi)^4}\,
v^{++}(-p,\theta,u_1) v^{++}(p,\theta,u_2) \qquad\nonumber\\
&& \times \Big(\widetilde I_{\mbox{\scriptsize
gauge}+\mbox{\scriptsize FP}} + \widetilde I_\varphi + \widetilde
I_\phi + O(\alpha_0)\Big).
\end{eqnarray}

\noindent The contributions of the matter and Pauli--Villars
superfields coincide with the corresponding contributions to Eq.
(\ref{Two_Point_V}), $\widetilde I_\phi = I_\phi$ and
$\widetilde I_\varphi = I_\varphi$. The remaining part of the result can be presented in the form

\begin{equation}
\widetilde I_{\mbox{\scriptsize gauge}+\mbox{\scriptsize FP}} +
\widetilde I_\varphi = -8\pi C_2 \int \frac{d^4q}{(2\pi)^4}
\frac{d}{d\ln\Lambda}
\Big(\frac{1}{q^4} - \frac{1}{(q^2+M_0^2)^2}\Big) = -\frac{C_2}{\pi}.\qquad
\end{equation}

\noindent
(Again, the terms containing higher derivatives are present at the intermediate steps of the calculation, but cancel each other in the final result.)
Therefore, the overall contributions to the two-point Green functions of the background and quantum gauge superfields are given by
the same integrals. This implies that all divergencies are
absorbed into the renormalization of the coupling constant and the
quantum gauge superfield is not renormalized, $Z_v=1+O(\alpha_0^2)$.

Thus, we see that the version of the higher covariant derivative
regularization proposed in this paper allows regularizing all
one-loop divergencies and subdivergencies. Moreover, using this regularization we have calculated all renormalization constants which encode all divergences of the considered theory.

\section{Summary}
\hspace{\parindent}

In this paper we formulate the higher covariant derivative
regularization and corresponding background field method for ${\cal
N}=2$ supersymmetric gauge theories in the harmonic superspace. This
regularization is completely mathematically consistent and does not
break the ${\cal N}=2$ supersymmetry and gauge invariance of the
theory in calculating the effective action. Using of ${\cal N}=2$
harmonic superspace allows to make the gauge fixing procedure in a
manifestly ${\cal N}=2$ supersymmetric way. Due to the background
field method the quantum corrections are also invariant under the
background gauge transformations. Thus, we construct the procedure
which allows to calculate loop quantum contributions to the effective
action without loss of manifest ${\cal N}=2$ supersymmetry and
gauge invariance. As a result, we justify an assumption in proof of
the ${\cal N}=2$ non-renormalization theorem implied in the
previous proof of this theorem. Also we illustrate application of
the constructed regularization by the explicit calculation of the one-loop renormalization constants
for the general renormalizable ${\cal N}=2$ SYM theory.

\section{Acknowledgments}
\hspace{\parindent}

The work of I.L.B and N.G.P is supported in parts by the RFBR grant,
project No 15-02-06670, grant for LRSS and project No 88.2014.2.
I.L.B. is grateful for DFG grant, project LE 8381/12-2 for partial
support. The research of I.L.B was supported by Ministry of
Education and Science of Russian Federation, project 3.867.2014/K.
The work of K.S. is supported by the RFBR grant, project No.
14-01-00695.


\end{document}